\documentclass[twocolumn]{aastex61}

\usepackage{amsmath,amssymb,times}
\usepackage[T1]{fontenc}  %% for \textquotedbl used in table

\usepackage{multirow}
\usepackage{url}
\usepackage{breakurl}

\shorttitle{Coronal Field Topology}

\shortauthors{DeRosa \& Barnes}

\def\refereeadd#1{{#1}}
\def\refereeout#1{}

\newcommand{\jplph}{J.\ Plasma Phys.}
\newcommand{\jasa}{J.\ Am.\ Stat.\ Assoc.}

\newcommand{\phpl}{Phys.\ Plasmas}
\newcommand{\rsta}{Phil.\ Trans.\ R.\ Soc.\ London, Ser.\ A}

\begin{document}

\title{Does Nearby Open Flux Affect the Eruptivity of Solar Active Regions?}

\author[0000-0002-6338-0691]{Marc~L.~DeRosa} \affil{Lockheed Martin Solar and
  Astrophysics Laboratory, 3251 Hanover St. B/252, Palo Alto, CA 94304, USA}

\author[0000-0003-3571-8728]{Graham~Barnes} \affil{NorthWest Research
  Associates, 3380 Mitchell Ln., Boulder, CO 80301, USA}

\begin{abstract}
The most energetic solar flares are typically associated with the ejection of
a cloud of coronal material into the heliosphere in the form of a coronal mass
ejection (CME). However, there exist large flares which are not accompanied by
a CME.  The existence of these non-eruptive flares raises the question of
whether such flares suffer from a lack of access to nearby open fields in the
vicinity above the flare (reconnection) site.  In this study, we use a sample
of 56 flares from Sunspot Cycles~23 and~24 to test whether active regions that
produce eruptive X-class flares are preferentially located near coronal
magnetic field domains that are open to the heliosphere, as inferred from a
potential field source surface model. The study shows that X-class flares
having access to open fields are eruptive at a higher rate than those for
which access is lacking. The significance of this result should be moderated
due to the small number of non-eruptive X-class flares in the sample, based on
the associated Bayes factor.
\end{abstract}


\section{Introduction}

The Sun is an active star that possesses a continually evolving, magnetized
corona. The continuous, quasi-steady evolution that is observed to occur most
of the time is occasionally interrupted by solar flares, which represent the
very rapid conversion of built-up magnetic free energy into light, heat, and
kinetic motions on time scales of about a few minutes.

The magnetic energy conversion occurs via magnetic reconnection, which is
assumed to take place in a thin layer of the corona where the Ohmic
resistivity is high enough to facilitate the transfer of energy from the field
to the plasma. Understanding the detailed dynamics of both the reconnection
region and how flares are triggered are areas of active research (see, e.g.,
the recent review by \citealt{jan2017} and references therein).

Theoretical treatments indicate that reconnection is more likely to occur at
particular locations in the magnetic field topology, such as null points,
separatrix surfaces, and quasi-separatrix layers (as reviewed by, e.g.,
\citealt{pon2012}). Determining the locations of such topologically important
features in the solar corona is challenging, however, owing to the lack of
direct measurements of the coronal magnetic field on the spatial scales needed
to discern these features. Although the coronal field topology is often more
evident from observations off the solar limb (e.g.,
\citealt{lin2004,gib2017}), the complementary photospheric magnetic field
observations needed for proper interpretation of the observed limb structures
are compromised by foreshortening.

In practice, indirect measurements of coronal magnetic field topology,
typically inferred using some combination of coronal imagery and coronal field
modeling, are used. Such indirect methods have been used to investigate a
broad range of properties, including the persistence of bright loop fans
surrounding active regions (ARs) that are otherwise quiescent \citep{sch2010};
how open flux maps down to the photosphere \citep{ant2011,pla2014}; cusps in
coronal limb observations \citep{fre2015}; flare ribbon geometries and
evolution \citep{zhao2014b,zhao2016,pon2016}; the temporal concurrence of
spatially separated events, including ``sympathetic flares''
\citep{sch2011,jin2016}; how the solar wind may be related to AR upflows in
regions of apparently closed fields \citep{edw2016}; and why the composition
of the solar wind near the boundaries between open and closed field appears to
be a mixture of plasma from both open and closed regions \citep{pon2015}.

Solar flares are most often characterized by their emission in X-ray
wavelengths, as detected by the X-ray spectrometers on board the various
\textit{Geostationary Orbiting Environmental Satellite} (GOES) missions over
the years, operated by the U.S.\ National Oceanic and Atmospheric
Administration (NOAA). The GOES flare catalog\footnote{At the time of this
  writing, yearly lists of GOES flares dating back to September~1975 can be
  downloaded at
  \url{https://www.ngdc.noaa.gov/stp/space-weather/solar-data/solar-features/solar-flares/x-rays/goes/xrs/}.}
categorizes flares in terms of their peak flux in the 1--8~\AA\ wavelength
band. The strongest and brightest flares, X-class flares, have a peak flux in
the 1--8~\AA\ band of at least 10$^{\text{--4}}$~W~m$^{\text{--2}}$ and are
often associated with coronal mass ejections (CMEs), in which a cloud of
coronal material is observed to be accelerated upward against gravity, away
from the Sun, and into the heliosphere. Although there is some correlation
between the X-ray emission of GOES flares and the properties and
characteristics (e.g., ejection velocities) of the ensuing CMEs, direct
proportionality should not be assumed \citep{ems2012}. It is thus important to
keep in mind that the peak X-ray emission from a flare is not necessarily a
good indicator of the total energy involved in the reconnection process.

Indeed, some X-class flares are not followed by any discernible eruption, such
as SOL2011-11-03T20:27 from NOAA~AR~11339 \citep{liu2014} and the cluster of
X-class flares from AR~12192 in October of 2014 \citep{sun2015}. An
understanding of why most X-class flares are accompanied by CMEs, but some are
not, probably depends on detailed knowledge of the forces responsible for the
upward acceleration of the coronal material at the core of the flaring AR, how
these forces compare to the downward forces that confine this material in the
lower corona, and the partitioning of energy resulting from the reconnection
process.

The scenario in which an X-class flare is not followed by any noticeable
eruption is intriguing. For the majority of X-class flares, the large amount
of energy associated with the X-ray emission is usually accompanied by enough
additional energy to overcome the confining forces and accelerate coronal
plasma into the heliosphere. The existence of non-eruptive X-class flares,
however, raises the question of whether such flares suffer from a lack of
access to nearby open fields above the flare (reconnection) site in which
overlying closed fields effectively block the upward rise of lower-lying flux
structures.  The observational and numerical studies by \citet{tor2017a} and
\citet{tor2017b} show that the ratio between the amount of flux involved in
the reconnection process and the total AR flux is smaller for non-eruptive
flares than for eruptive flares, supporting this possibility. A related study
by \citet{wan2017} also indicates a propensity for the large-scale coronal
field associated with non-eruptive flaring ARs to be more confining.

If the hypothesis presented above is true, then one would expect a decreased
likelihood of eruptivity in cases where the flaring AR is buried more deeply
underneath a significant amount of closed magnetic fields. Stated more
broadly, identifying whether the presence or absence of particular topological
features in the large-scale coronal magnetic field is correlated with whether
a flare is confined or eruptive may be a useful diagnostic of the propensity
of a flaring AR to foster an eruption.

In the study presented here, we investigate whether the nature of the coronal
fields that lie above the locations of strong flares is a contributing factor
in determining whether these flares are accompanied by plasma ejected into the
heliosphere. More specifically, we test the hypothesis that ARs in which
eruptive flares occur are preferentially located near open fields, and
conversely that ARs in which confined flares occur are preferentially located
underneath closed topological structures. To perform this test, we apply
topological analysis software to models of the global coronal magnetic field
corresponding to the times of 56 X-class flares in the GOES flare catalog from
the past two decades spanning sunspot Cycles~23 and~24. Using statistical
methods, we estimate the rate at which flares from ARs with access to open
field are eruptive and compare this estimate to the rate from ARs under closed
field.

\section{Methodology}

\subsection{Obtaining the Flare Sample}

According to the GOES database, 176 X-class flares have occurred since the
beginning of sunspot Cycle~23 in~1996.\footnote{The following query to the
  Heliophysics Events Knowledgebase \citep{hur2012} yields the full list of
  GOES X-class flares occurring between 1996 and 2017:
  \url{https://www.lmsal.com/isolsearch?hek_query=https://www.lmsal.com/hek/her?cosec=2&&cmd=search&type=column&event_type=fl&event_starttime=1996-01-01&event_endtime=2018-01-01&event_region=all&event_coordsys=helioprojective&x1=-5000&x2=5000&y1=-5000&y2=5000&result_limit=200&sparam0=FL_GOESCls&op0=\%3E=&value0=X1&sparam1=FRM_Name&op1==&value1=SWPC}}
The large-scale magnetic environment surrounding each flaring AR may be
assessed using models of the global solar coronal magnetic field, including
the oft-used potential field source-surface (PFSS) model used in this
study.

Determining the magnetic environment associated with each flare location
presupposes that the flare location is known. However, some X-class flares in
the GOES database from sunspot Cycles~23 and~24 have indeterminate locations,
which unfortunately results in their removal from the sample unless the
location of the AR can be determined by other means. During sunspot Cycle~23,
the absence of H-alpha images contemporaneous with the flares is a significant
factor in the lack of locational knowledge. During sunspot Cycle~24 all flares
on disk can be determined using the frequent imagery from the Atmospheric
Imaging Assembly (AIA; \citealt{lem2012}) instrument on the Solar Dynamics
Observatory (SDO). Flares at the limb suffer from the issue of geometrical
foreshortening that makes determinations of precise longitudes difficult.

The PFSS approximation assumes that the coronal volume is current-free,
enabling the magnetic field within a spherical shell to be calculated given
full-Sun magnetic maps of the photosphere \citep{sch1969}. The boundary
conditions are completely specified if it is also assumed that the magnetic
field is purely radial at the upper boundary. In this study, the lower
boundary at $R_\text{bot}=R_\odot$ in the PFSS models are provided by sampling
the evolving flux-transport model of \citet{sch2003}, in which magnetograms
from either the Michelson Doppler Imager (MDI; \citealt{sch1995}) on board the
Solar and Heliospheric Observatory (SOHO) spacecraft (between 1996 and 2010)
or the Helioseismic and Magnetic Imager (HMI; \citealt{sch2012}) on board SDO
(after 2010) are incorporated into the model. The radius of the upper boundary
is chosen to be the canonical value of $R_\text{top}=2.5 R_\odot$.  Both such
models used here, namely the evolving surface-flux models of the photospheric
magnetic field and the subsequent PFSS models of the coronal magnetic field,
are publicly available for download via the \texttt{pfss}\ package from the
SolarSoftWare (SSW) distribution system.

PFSS models are affected by the additional issue that new flux is only
incorporated into the model after it appears in MDI or HMI
magnetograms. Occasionally, an AR located at or near the east limb that
contains a significant amount of flux and that is not yet incorporated into
the model does affect the arrangement of coronal magnetic fields at on-disk
longitudes (as explored in, e.g., \citealt{nit2008} or \citealt{sch2011}). In
the more extreme cases, unaccounted east-limb flux may affect the global field
situated as far away as 90$^\circ$ to 120$^\circ$ of longitude. We therefore
have more confidence in the modeled magnetic fields for locations west of the
central meridian (i.e., farther away from possible missing flux on the east
limb) than for locations in the eastern hemisphere, and as a result we have
screened out all flares with locations east of the central meridian. The final
sample comprises 56 flares occurring within 37 ARs, as detailed in
Table~\ref{table:flaretable}.

\subsection{PFSS Model Sanity Checks}

PFSS models assume a current-free magnetic field solution, and thus these
models are not physically appropriate in the low coronae in the cores of ARs,
where significant currents are known to exist. However, farther away from ARs,
PFSS models often possess field geometries that resemble many larger-scale
features observed in the solar coronal magnetic field, suggesting that much of
the coronal volume is largely current-free. Because this investigation
considers only coronal magnetic fields on larger spatial scales, PFSS models
are assumed appropriate; it nonetheless seems prudent to evaluate the
resemblance between observations and the PFSS models for the specific times
considered here to see whether there are any significant discrepancies, as a
sanity check.

To this end, we employ two qualitative tests: (1) comparisons between the
topological structures found in PFSS models with the locations of streamers
and pseudostreamers evident in white-light coronagraph images, and (2)
comparisons between the locations of coronal holes visible in extreme
ultraviolet (EUV) images with the open-flux regions determined from the PFSS
models. For all events, these tests either support the idea that the coronal
magnetic field is current-free on large scales, or were inconclusive. The
online materials associated with this article
(\url{http://www.lmsal.com/forecast/DB2018.html}) provide images and
topological renderings for each of the 56 events used for this
investigation. These images allow the reader to assess the applicability of
the PFSS model in the manner described in this section.

Although more rigorous comparison schemes are possible, these involve more
physically realistic modeling of the coronal magnetic field. These more
rigorous tests are not considered here, as such modeling requires knowledge of
(at least) photospheric currents, plasma densities and temperatures, and/or
coronal heating mechanisms --- quantities that are generally not readily
available for a large enough sample of ARs and for a large enough area on the
Sun. Additionally, these models are more computationally intensive and are not
as readily applied to a large sample of regions.

\begin{figure*}
  \begin{center}
  \includegraphics[width=\textwidth]{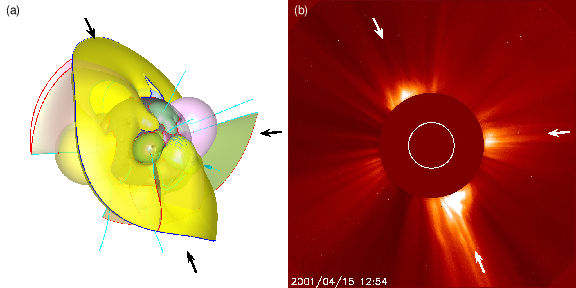}
  \caption{SOL2001-04-15T13:50 (Event~11 in Table~\ref{table:flaretable})
    comparison: (a) Topological skeleton for the PFSS model of 2001 April~15
    at 12:04~UT and (b) the corresponding LASCO~C2 image. The topological
    skeleton comprises the separatrix surfaces (semi-transparent surfaces),
    null points (small red dots), and spine lines (cyan lines) present in the
    PFSS model. The dark-blue line is the polarity-inversion line at the upper
    boundary of the model at $R_\text{top}=2.5 R_\odot$. Red lines are drawn
    where the separatrix curtains intersect $R_\text{bot}$ and
    $R_\text{top}$. The arrows in both images indicate the positions of the
    brightest LASCO streamers, which correspond to the largest separatrix
    surfaces in the topology skeleton. The conical cyan pointer indicates the
    location of the X-class flare at S20W85.}
  \end{center}
  \label{fig:skeleton_lasco-1}
\end{figure*}

\begin{figure*}
  \begin{center}
  \includegraphics[width=\textwidth]{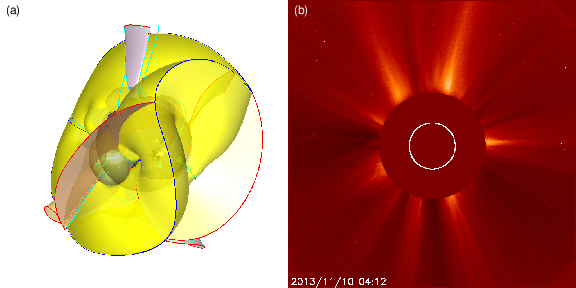}
  \caption{Comparison between the (a) topological skeleton and (b) LASCO~C2
    image as in Figure~\ref{fig:skeleton_lasco-1}, but for SOL2013-11-10T05:14
    (Event~48 in Table~\ref{table:flaretable}). The conical cyan pointer
    indicating the X-class flare location S14W13 is visible (barely)
    underneath the HCS curtain. Here, there are several LASCO streamers and it
    is more challenging to associate their positions with features in the PFSS
    model topology. Likewise, it is hard to anticipate the locations of LASCO
    streamers from the PFSS topology skeleton.}
  \end{center}
  \label{fig:skeleton_lasco-2}
\end{figure*}

\subsubsection{Comparisons with (Pseudo-)Streamers}

The first sanity check is based on the fact that the cusp-shaped streamers
evident in white-light coronagraph images from, e.g., the Large Angle and
Spectrometric Coronagraph (LASCO; \citealt{bru1995b}) on SOHO result from the
increased density associated with tall, high-arching closed fields that
underlie the heliospheric current sheet (HCS). If a PFSS model successfully
captures the largest spatial scales in the actual coronal field, then the
position angles of the LASCO streamers are expected to correspond with the
highest-arching closed-field structures in the PFSS model. The persistence of
coronal streamers over several rotation periods (an observational fact that
was first realized approximately 50 years ago, e.g., \citealt{boh1970a}) lends
credence that streamers are a robust feature of the large-scale corona. The
magnetic null points associated with these streamers located lower down in the
corona have also been found to persist \citep{fre2015}.

Examples of this first test are demonstrated in
Figures~\ref{fig:skeleton_lasco-1} and~\ref{fig:skeleton_lasco-2},
corresponding to the times of Events~11 and~48 in
Table~\ref{table:flaretable}. In both figures, the topological skeleton
associated with the PFSS model nearest to the time of the event is shown in
panel~(a) and a corresponding LASCO~C2 image in panel~(b). The topological
skeleton renderings shown here are largely similar to those shown in
\citet{pla2014}, and illustrate the separatrix surfaces, null points, and
spine lines in the PFSS models. These features are depicted in the figures as
semi-transparent surfaces, small red dots, and cyan lines, respectively. We
note as an aside that the same topological elements of interest found in the
PFSS models, such as the location of null points and the boundaries between
magnetic field connectivity domains, are likely to also be present in
non-potential fields \citep{reg2012}. The algorithms by which the topological
features were calculated are the null-point finding method of \citet{hay2007}
and the separatrix-surface mapping scheme described in \citet{hay2010}, after
adapting for spherical geometries.

In the comparison with LASCO images, the most relevant topological features in
the PFSS models are the separatrix surfaces that intersect $R_\text{top}$. The
largest and most noticeable separatrix surfaces of this kind are the surfaces
that extend downward from the polarity-inversion line at $R_\text{top}$, and
serve to separate fieldlines that are considered open to the heliosphere
(i.e., fieldlines that have one endpoint at $R_\text{bot}$ and another at
$R_\text{top}$) from closed fieldlines (i.e., fieldlines having both endpoints
at $R_\text{bot}$). These \textsl{HCS curtains} (as termed by
\citealt{pla2014}) are colored yellow in Figures~\ref{fig:skeleton_lasco-1}(a)
and~\ref{fig:skeleton_lasco-2}(a), and the polarity-inversion line at
$R_\text{top}$ at the apexes of these surfaces is colored dark blue. HCS
curtains are conceptualized to continue upward to form the HCS (as in
  Figure~1 of their paper) and to be at the same position angles in the PFSS
  models, when viewed from along the Earth-Sun line, as the white-light
  coronal streamers seen in LASCO images \citep{wan2007b}.

Additionally, each null point located inside PFSS coronal volume has a
separatrix surface associated with it. While these surfaces often take the the
shape of domes that wall off self-contained domains of fieldlines covering
sections of the photosphere, in some cases the fan plane extending away from a
null point is found to be oriented vertically, such that the associated
separatrix surface extends upward and intersects $R_\text{top}$. These
\textsl{separatrix curtains} (as termed by \citealt{pla2014}) divide open
fieldlines having the same polarity, and are often associated with coronal
pseudostreamers observed in the LASCO images \citep{wan2007a}. In
Figures~\ref{fig:skeleton_lasco-1}(a) and~\ref{fig:skeleton_lasco-2}(a), all
separatrix surfaces associated with coronal nulls (including the separatrix
curtains) are rendered in various pastel colors. The intersection of these
separatrix surfaces with either the upper or lower boundary are colored red.

Because the HCS curtains and separatrix curtains are both associated with
LASCO streamers and pseudostreamers, comparisons between the renderings of the
topological skeletons of the PFSS models (centered on the solar
central-meridian longitude and latitude for the time of interest) and the
LASCO images provide a way to validate the PFSS models. In
Figure~\ref{fig:skeleton_lasco-1}, corresponding to Event~11, the position
angles of the three brightest streamers in LASCO (marked by arrows) match well
with the HCS curtains and one of the upward-extending separatrix curtains. As
a result, the PFSS model for Event~11 is considered plausible.

In Figure~\ref{fig:skeleton_lasco-2}, corresponding to Event~48, the
comparison is less conclusive. The LASCO image contains a multitude of
streamers and pseudostreamers. The PFSS model topology is also more complex,
with an undulating and warped HCS curtain surrounded by many smaller
separatrix curtains. In this case, it is more difficult to predict where
streamers might occur by looking only at the topological rendering, and it is
correspondingly difficult to choose features in the topological rendering that
match the LASCO streamers. Streamers and pseudostreamers are only evident when
there is a significant amount of plasma density along the line of sight, and
this suggests that the orientation of the HCS curtains may affect the presence
or absence of streamers, especially if the HCS curtains are more face-on than
edge-on. Figure~\ref{fig:skeleton_lasco-2}(a) indicates that the HCS curtains
for Event~48 are more folded and undulated, with portions being oriented
face-on. As a consequence, for this particular case the comparison is deemed
inconclusive.

\begin{figure*}
  \begin{center}
    \includegraphics[width=\textwidth]{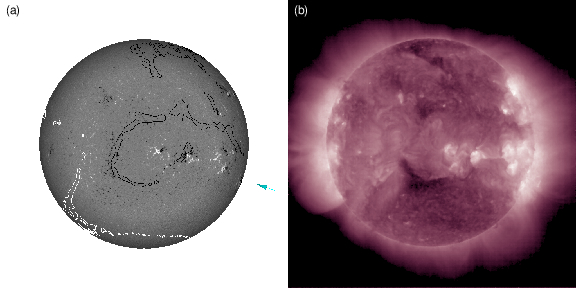}
    \caption{SOL2001-04-15T13:50 (Event~11 in Table~\ref{table:flaretable})
      comparison: (a) Outlines of open flux from the PFSS model of 2001
      April~15 at 12:04~UT overlaid on the nearest MDI line-of-sight
      magnetogram and (b) the corresponding full-Sun image from the
      284\AA\ channel of EIT. The conical cyan pointer in panel~(a) indicates
      the location of the X-class flare at S20W85. The open-flux contours from
      the PFSS model match the coronal holes observed by EIT reasonably well,
      in this case.}
  \end{center}
  \label{fig:opflux_euv-1}
\end{figure*}

\begin{figure*}
  \begin{center}
    \includegraphics[width=\textwidth]{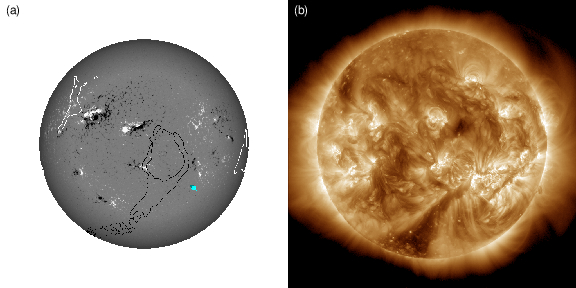}
    \caption{As in Figure~\ref{fig:opflux_euv-1}, a comparison for
      SOL2014-12-20T00:28 (Event~52 in Table~\ref{table:flaretable}) between
      the (a) outlines of open flux from the PFSS model and (b) the
      corresponding full-Sun image from the 193\AA\ channel of AIA. The
      conical cyan pointer in panel~(a) indicates the location of the X-class
      flare at S21W24. The open-flux contours from the PFSS model match the
      coronal holes observed by AIA reasonably well, in this case.}
  \end{center}
  \label{fig:opflux_euv-2}
\end{figure*}

\subsubsection{Comparisons with Coronal Holes}

The second sanity check relies on the association of dark regions in EUV
images with open fields. The plasma along open fieldlines is too cool and too
rarefied to emit in EUV wavelengths, and much of it is instead streaming
upward to become the solar wind. Therefore, comparisons between the open-flux
domains predicted by the PFSS model and the dark regions in EUV and X-ray
imagery can be used to gauge how well the large-scale coronal magnetic field
is represented by the PFSS models.

Such comparisons are imperfect, and a one-to-one correspondence between
EUV-dark regions in the images and open-flux domains from the model is not
expected (see \citealt{low2014,low2017} for recent comparisons). Several
reasons probably account for these discrepancies: (1) Coronal holes may not
indicate open flux; instead, this plasma may be located on long, closed
fieldlines that connect faraway regions of opposite polarities. The plasma
found on such long fieldlines is often not at the proper density or
temperature to emit in EUV wavelengths, and thus remains dark. (2) The corona
is optically thin, and as a result lines of sight passing through both closed
and open fields will almost always appear bright. Such bright coronal
structures may obscure open-field channels, especially away from disk center,
and as a result open-flux regions will not appear dark in the EUV if there is
not a direct line of sight into the channel. (3) The static upper boundary of
the PFSS model only crudely approximates the dynamic environment present at
the boundary between the magnetism-dominated corona and the plasma-dominated
heliosphere. One consequence of this situation is that measurements of the
{\em in situ} open flux at 1~AU do not match that predicted by the PFSS model
\citep{lin2017}. (4) The PFSS open-flux domains are large-scale features that
span the full height of the model, and as a result may be affected by the lack
of up-to-date surface magnetic fields at east-limb longitudes (as in the case
discussed in \citealt{pev2016}). Structures in the eastern hemisphere may be
adversely affected when there is a significant amount of flux at or past the
east limb that has not yet been assimilated into the surface-flux model that
comprises the lower-boundary condition of the PFSS extrapolation.

Figure~\ref{fig:opflux_euv-1} illustrates the comparison between the modeled
open flux and coronal holes for Event~11 (the same event shown in
Figure~\ref{fig:skeleton_lasco-1}). The image in
Figure~\ref{fig:opflux_euv-1}(a) shows a magnetogram for 2001 April~15 at
about 0~UT, on which are overplotted the outlines of the photospheric
footpoints of field lines that intersect $R_\text{top}$. Open fieldlines fan
out from these contours, sometimes with significant expansion factors.  The
colors of the open-field contours in the figure indicate the polarity of the
open flux. These open-field contours may be qualitatively compared with the
darker regions of the full-Sun image from the 284\AA\ channel observed by the
Extreme ultraviolet Imaging Telescope (EIT; \citealt{del1995}) on SOHO, shown
in Figure~\ref{fig:opflux_euv-1}(b).

In the case shown in Figure~\ref{fig:opflux_euv-1}, there is a fair amount of
correspondence between many of the open-flux contours and the coronal holes,
with the shapes of the darker features in the EIT image bearing resemblance to
several of the shapes of the open-flux contours. In particular, the coronal
hole and the PFSS open flux domain at the north pole have similar
outlines. Similarly, the curved shape of open-flux region near the central
meridian that spans the near-equatorial latitudes resembles the corresponding
darker channels in the EUV image, though the degree to which they match is not
as good for this equatorial coronal hole as in the polar coronal hole
described earlier. This region of open flux is narrow, and brighter plasma
associated with neighboring closed fields may be obscuring the full coronal
hole. The same effect may be why the modeled open-flux extension in the
southeast quadrant of Figure~\ref{fig:opflux_euv-1}(a) has no noticeable
coronal hole in the EUV image of Figure~\ref{fig:opflux_euv-1}(b), although
east-limb open-flux contours may also be affected by inaccuracies in the
photospheric boundary condition.

Figure~\ref{fig:opflux_euv-2} shows the same comparison for Event~52. On this
date, a PFSS open-flux region extends northward from the south polar
region. At the top of this extension there is a ring of open flux that
surrounds a closed-field domain. The comparison image from the 193\AA\ channel
of AIA also contains a dark coronal hole extending in the same direction as in
the model, as well as some evidence that a circular channel of open flux might
be present. Additionally, the coronal hole in the northeastern quadrant of the
AIA image that appears to extend behind the limb matches well with the
location of an open-flux region evident in the same location in the PFSS
model.

\setlength{\extrarowheight}{-0.12em}
\startlongtable
\begin{deluxetable*}{cccccccc}
\tablecaption{Sample of flaring active regions\label{table:flaretable}}
  \tablehead{\colhead{Event} & \colhead{SOL (flare peak time)\tablenotemark{a}} & \colhead{flare class\tablenotemark{b}} & \colhead{NOAA AR\tablenotemark{c}} & \colhead{Location\tablenotemark{d}} & \colhead{Eruptive?\tablenotemark{e}} & \colhead{Access?\tablenotemark{f}} & \colhead{Notes} }
  \startdata
1 & SOL1996-07-09T09:12 & X2.6 & 7978 & S10W30 & Yes & No & \\
2 & SOL1997-11-04T05:58 & X2.1 & 8100 & S14W33 & Yes & Yes & \\
3 & SOL1998-05-02T13:42 & X1.1 & 8210 & S15W15 & Yes & Yes & $g$\\
4 & SOL1999-08-28T18:05 & X1.1 & 8674 & S26W14 & Yes & No & $h$\\
5 & SOL1999-11-27T12:12 & X1.4 & 8771 & S15W68 & No & No & \\
6 & SOL2000-11-24T15:13 & X2.3 & 9236 & N22W7 & Yes & Yes & \\
7 & SOL2000-11-25T18:44 & X1.9 & \textquotedbl & N20W23 & Yes & Yes & \\
8 & SOL2000-11-26T16:48 & X4.0 & \textquotedbl & N18W38 & Yes & Yes & \\
9 & SOL2001-03-29T10:15 & X1.7 & 9402 & N20W19 & Yes & Yes & $g$\\
10 & SOL2001-04-10T05:26 & X2.3 & 9415 & S23W9 & Yes & Yes & \\
11 & SOL2001-04-15T13:50 & X14. & \textquotedbl & S20W85 & Yes & Yes & \\
12 & SOL2001-10-19T16:30 & X1.6 & 9661 & N15W29 & Yes & No & $h$\\
13 & SOL2001-10-25T15:02 & X1.3 & 9672 & S16W21 & Yes & No & $h$\\
14 & SOL2001-11-04T16:20 & X1.0 & 9684 & N6W18 & Yes & No & \\
15 & SOL2002-07-18T07:44 & X1.8 & 10030 & N19W30 & Yes & Yes & \\
16 & SOL2002-08-21T05:34 & X1.0 & 10069 & S12W51 & Yes & Yes & $i$\\
17 & SOL2003-03-17T19:05 & X1.5 & 10314 & S14W39 & Yes & No & \\
18 & SOL2003-03-18T12:08 & X1.5 & \textquotedbl & S15W46 & Yes & No & $j$\\
19 & SOL2003-05-27T23:07 & X1.3 & 10365 & S7W17 & Yes & No & \\
20 & SOL2003-05-29T01:05 & X1.2 & \textquotedbl & S6W37 & Yes & No & \\
21 & SOL2003-10-26T18:19 & X1.2 & 10484 & N2W38 & Yes & Yes & \\
22 & SOL2003-10-29T20:49 & X10. & 10486 & S15W2 & Yes & Yes & \\
23 & SOL2003-11-02T17:25 & X8.3 & \textquotedbl & S14W56 & Yes & Yes & \\
24 & SOL2003-11-03T01:30 & X2.7 & 10488 & N10W83 & Yes & No & \\
25 & SOL2003-11-03T09:55 & X3.9 & \textquotedbl & N8W77 & Yes & No & \\
26 & SOL2003-11-04T19:50 & X28. & 10486 & S19W83 & Yes & Yes & \\
27 & SOL2004-02-26T02:03 & X1.1 & 10564 & N14W15 & No & No & $j$\\
28 & SOL2004-08-13T18:12 & X1.0 & 10656 & S13W24 & Yes & Yes & \\
29 & SOL2004-08-18T17:40 & X1.8 & \textquotedbl & S14W90 & Yes & Yes & \\
30 & SOL2004-10-30T11:46 & X1.2 & 10691 & N13W25 & Yes & Yes & \\
31 & SOL2004-11-07T16:06 & X2.0 & 10696 & N9W17 & Yes & No & $j$\\
32 & SOL2004-11-10T02:13 & X2.5 & \textquotedbl & N9W49 & Yes & No & \\
33 & SOL2005-01-15T23:02 & X2.6 & 10720 & N14W8 & Yes & No & \\
34 & SOL2005-01-17T09:52 & X3.8 & \textquotedbl & N15W25 & Yes & Yes & \\
35 & SOL2005-01-19T08:22 & X1.3 & \textquotedbl & N15W51 & Yes & Yes & \\
36 & SOL2005-01-20T07:01 & X7.1 & \textquotedbl & N14W61 & Yes & Yes & \\
37 & SOL2005-07-14T10:55 & X1.2 & 10786 & N11W90 & Yes & No & $h$\\
38 & SOL2005-09-15T08:38 & X1.1 & 10808 & S12W14 & No & Yes & $i$\\
39 & SOL2006-12-13T02:40 & X3.4 & 10930 & S6W23 & Yes & Yes & \\
40 & SOL2006-12-14T22:15 & X1.5 & \textquotedbl & S6W46 & Yes & Yes & \\
41 & SOL2011-02-15T01:56 & X2.2 & 11158 & S20W10 & Yes & No & $j$\\
42 & SOL2011-03-09T23:23 & X1.5 & 11166 & N8W11 & No & No & $h$\\
43 & SOL2011-08-09T08:05 & X6.9 & 11263 & N14W69 & Yes & No & \\
44 & SOL2011-09-06T22:20 & X2.1 & 11283 & N14W18 & Yes & Yes & \\
45 & SOL2011-09-07T22:38 & X1.8 & \textquotedbl & N14W31 & Yes & Yes & \\
46 & SOL2012-07-12T16:49 & X1.4 & 11520 & S13W3 & Yes & No & $h$\\
47 & SOL2013-10-28T02:03 & X1.0 & 11875 & N4W66 & Yes & Yes & \\
48 & SOL2013-11-10T05:14 & X1.1 & 11890 & S14W13 & Yes & No & $h$\\
49 & SOL2014-03-29T17:48 & X1.0 & 12017 & N10W32 & Yes & Yes & $i$\\
50 & SOL2014-10-26T10:56 & X2.0 & 12192 & S14W37 & No & No & $h$\\
51 & SOL2014-10-27T14:47 & X2.0 & \textquotedbl & S16W56 & No & No & $h$\\
52 & SOL2014-12-20T00:28 & X1.8 & 12242 & S21W24 & Yes & Yes & \\
53 & SOL2017-09-06T09:10 & X2.2 & 12673 & S8W32 & Yes & Yes & $g$\\
54 & SOL2017-09-06T12:02 & X9.3 & \textquotedbl & S9W34 & Yes & Yes & $g$\\
55 & SOL2017-09-07T14:36 & X1.3 & \textquotedbl & S8W48 & Yes & Yes & \\
56 & SOL2017-09-10T16:06 & X8.2 & \textquotedbl & S8W88 & Yes & Yes & \\
  \enddata
  \tablenotetext{a}{Solar Object Locator (SOL) of time of peak flare emission from the \textit{GOES} flare catalog}
  \tablenotetext{b}{Flare class from the \textit{GOES} flare catalog}
  \tablenotetext{c}{Active region number assigned by NOAA}
  \tablenotetext{d}{Flare location from the \textit{GOES} flare catalog}
  \tablenotetext{e}{Is there an eruption in LASCO C2 and/or C3 data following the time of the flare peak?}
  \tablenotetext{f}{Does the PFSS model imply access to open field from an upward-directed eruption centered on the flare location?}
  \tablenotetext{g}{Access to open fields is provided via a narrow channel located between separatrix surfaces. This channel extends either into or through the AR and encompasses the flare site, as in the example shown in Figure~\ref{fig:narrow-channel}.}
  \tablenotetext{h}{There is a significant volume of closed field above the flare site that likely blocks access to open fields for any flux structure that may accelerate upward, even though the flare location is laterally adjacent to open flux. An example of this phenomenon is shown in Figure~\ref{fig:skeleton_lasco-2}.}
  \tablenotetext{i}{The flare is sited near a (small, often) region of open field that significantly expands with height, creating a funnel- or fan-shaped open-flux domain that overlies any upwardly mobile flux structure located at the flare site. An example of this phenomenon is shown in Figure~\ref{fig:funnel-figure}.}
  \tablenotetext{j}{The location of the flare is underneath a separatrix dome associated with a null point located in the coronal volume, according to the PFSS model.}
\end{deluxetable*}
  %% table is generated from validcat.sav

\begin{figure*}
  \begin{center}
    \includegraphics[width=\textwidth]{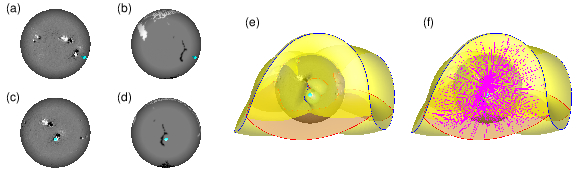}
    \caption{A narrow channel of open flux is associated with
      SOL2017-09-06T09:10 and SOL2017-09-06T12:02 (Events~53 and~54 in
      Table~\ref{table:flaretable}). Panels~(a) and~(b) show the photospheric
      magnetic fields and the photospheric open-flux regions from the PFSS
      extrapolation, respectively, as viewed from the Earth-Sun line, with
      color indicative of polarity. Panels~(c) and~(d) are the same as
      panels~(a) and~(b), except that the models have been rotated so that the
      flare longitude is centered. Panels~(e) and~(f) show the topological
      separatrix surfaces as rendered on the magnetic map shown in panel~(c),
      with the HCS curtain (colored yellow) dominating the image. A separatrix
      curtain (colored orange) in the southern hemisphere divides the negative
      open flux into separate domains. Near disk center lies a narrow channel
      of open-flux that passes close to the flaring AR, as seen in panels~(d)
      and~(e). Field lines emanating from this particular domain, shown in
      panel~(f), are seen to have high expansion factors. In all panels, the
      conical cyan pointer indicates the location of the X-class flares.}
  \end{center}
  \label{fig:narrow-channel}
\end{figure*}

\begin{figure*}
  \begin{center}
    \includegraphics[width=\textwidth]{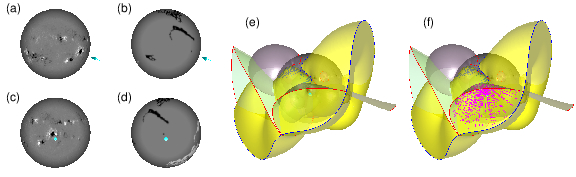}
    \caption{A small spot of open flux is located near SOL2002-08-21T05:34
      (Event~16 in Table~\ref{table:flaretable}). As in
      Figure~\ref{fig:narrow-channel}, panels~(a)--(d) show the photospheric
      magnetic fields and the photospheric open-flux regions from different
      perspectives, and panels~(e) and~(f) show the topological separatrix
      surfaces from the coronal field model. In all panels, the conical cyan
      pointer indicates the location of the X-class flare. A small spot of
      open flux is located in the trailing polarity of the AR associated with
      the event, visible in panel~(d) as the small blck region immediately
      northeast of the conical pointer. The modeled open flux expands outward
      and extends upward from this spot to occupy the volume between one of
      the separatrix curtains and the yellow HCS curtain. }
  \end{center}
  \label{fig:funnel-figure}
\end{figure*}

\section{Results and Discussion}

The final sample of events comprises 56 flares occurring within 37 ARs, as
listed in Table~\ref{table:flaretable} and in the online materials associated
with this article (\url{http://www.lmsal.com/forecast/DB2018.html}). Each
flare in the sample is classified as either eruptive or non-eruptive, based on
whether a CME is observed in LASCO data. We made use of the LASCO CME
catalog\footnote{At the time of this writing, the LASCO CME catalog can be
  found at \url{https://cdaw.gsfc.nasa.gov/CME_list/index.html}.} to determine
whether the flares have an associated CME. Examining LASCO C2 and C3 running
difference movies is particularly helpful for this purpose, and the LASCO CME
catalog has conveniently provided a useful movie-making tool that synchronizes
LASCO C2 and C3 running difference movies with GOES X-ray light curves. In the
online materials, clickable links to such synchronized movies are provided for
all 56 events.

We also characterize each event based on whether there is access to open
fields from the location of the flare. More specifically, we consider in a
qualitative manner how likely it is that a rising flux structure located at
the flare site would encounter open fields as it moves radially outward
through the PFSS model. In some cases, this is easily judged as, for example,
when the source AR is permeated by open fields, or when the source AR is
centered underneath the helmet surface (and is thus obviously buried beneath a
significant amount of closed field). Many cases are more ambiguous, and thus
making the determination is more subjective.

Narrow channels of open flux are a common feature in PFSS coronal field
models, and are usually nestled either between separatrix surfaces that divide
different topological domains of the magnetic field or between tight folds in
the HCS curtain. The fields emanating from these channels often have high
expansion factors, especially in the direction perpendicular to the channel
orientation, and are believed to play a key role in the formation of the slow
solar wind \citep{ant2007,ant2011,tit2011,hig2017b}. In the context of this
study, narrow channels that pass in or through a flaring AR provide a pathway
by which plasma and fields may be readily ejected into interplanetary space,
even though the flare site may not be located precisely above the photospheric
open-field footprint. Because of their small photospheric area, such
open-field channels are sometimes difficult to identify in EUV imagery.

As an example of this phenomenon, Figure~\ref{fig:narrow-channel} illustrates
a narrow channel encroaching upon the trailing polarity of AR~12673, which
produced the series of recent X-class flares in September 2017. Although the
closest region of photospheric open flux is not directly underneath the flare
site, we consider this region to have access to open fields because of how
quickly with height this nearby open flux splays out. A variant of this effect
involves open-flux domains with even smaller photospheric areas that map down
to strong flux, such as for SOL2002-08-21T05:34 shown in
Figure~\ref{fig:funnel-figure}. As with AR~12673, the open fields above
AR~10069 map down to a small, isolated region on the photosphere in the
trailing polarity of the flaring AR.

\begin{deluxetable}{cc|cc}
\tablecolumns{3}
\tablewidth{0pc}
\tablecaption{Contingency Table\label{table:contingency}}
\tablehead{
  \multirow{2}{*}{$n_\text{total}=56$} & & \multicolumn{2}{c}{Eruptive?} \\
  & & {\it Yes} & {\it No}} 
\startdata
\multirow{2}{*}{Access?} & {\it Yes}  & $n_{eo}=30$  & $n_{no}=1$ \\
& {\it No}   & $n_{ec}=20$  & $n_{nc}=5$ \\
\enddata
\end{deluxetable}

\refereeadd{Determining whether open flux can be associated with a flare is
  often more ambiguous. In the online materials for SOL2005-07-14T10:55
  (Event~37 in Table~\ref{table:flaretable}), it is evident that the flare
  location occurs near the southern extent of an open-flux region that
  stretches southward from the north pole.  Although the photospheric location
  of open flux extends very close to the latitude and longitude of the flare
  site, the topological domain of connectivity above the flare site contains a
  large volume of closed field that bows outward above the active
  region. Because eruptions are directed outward and upward, we judge in this
  case that the significant amount of overlying closed field would serve to
  confine any upward motion. As a result, this case and others like it are
  listed in Table~\ref{table:flaretable} as not having access to open fields
  due to the particular geometry of the closed-field domain located above the
  flare site. The series of X-class flares originating from AR~12192 also
  possess this property.}

Table~\ref{table:contingency} is a contingency table that summarizes the
number of events that fall into each of the defined categories. The tabulation
shows that of the 50 X-class flares associated with a CME, 30 of these (60\%)
occurred in locations judged as having access to open flux. There are only 6
non-eruptive X-class flares in the sample, and 5 of these were sited in places
with significant overlying closed fields. We estimate the rate at which
X-class flares with access to open flux are eruptive as 0.97 (30/31) compared with
0.80 (20/25) for X-class flares without access to open flux.  These estimates
are, however, based on a small number (6) of non-eruptive flares.

To test how robust the results are, we computed the Bayes factor $K$
\citep[e.g.,][]{kas1995} comparing a model in which the rate at which X-class
flares are eruptive is independent of access to open field with a model in
which access to open field results in a different rate of eruptions (see
Appendix~\ref{section:stats}). Depending on the choice of priors, the Bayes
factor is in the range $0.12 \le K \le 0.74$, which indicates that there is
weak to moderate evidence to support that access to open field influences
whether an X-flare is eruptive.

\section{Conclusions}

We conclude that X-class flares are more likely to be eruptive when they occur
in locations with access to open flux. The evidence to support this, however,
is statistically sensitive to the small number of non-eruptive X-class flares
in the sample. Of the 31 X-class flares that were judged to originate in
locations with access to open field, all except one (SOL2005-09-15T08:38) were
eruptive. The sample also contains 25 X-class flares located far away from
open fields, of which 20 were eruptive and 5 were non-eruptive. Access to open
field is therefore neither a necessary nor a sufficient condition for a flare
to result in an eruption.

\refereeadd{That proximity to open field is not a more clear-cut discriminator
  is an indication that other properties of the source AR also contribute to
  whether an X-class flare is associated with an eruption. For eruptive ARs,
  features such as the reconnection flux and the decay index have been
  demonstrated to be correlated with CME speeds
  \citep[e.g.,][]{liu2008,kaz2017,den2017}. Even though non-eruptive flares
  are not considered in these studies, we speculate that these aforementioned
  trends extend into the realm of non-eruptive flares, i.e., we suspect that
  flaring ARs without discernable eruptions involve less reconnected flux and
  a lower decay index than eruptive ARs, though we acknowledge that these
  trends should be established more rigorously using samples that include both
  eruptive and non-eruptive flares. Other source-region properties, such as
  the distance between the center of the AR and the flare site, may also be
  important \citep{wan2007c}.}

\refereeadd{The topology of the magnetic field ovelying a flare site is also
  thought to affect the chances of an eruption. For instance, the presence of
  a null point in the magnetic field may be necessary for an eruption to
  proceed or may otherwise facilitate an eruption
  \citep[e.g.,][]{dem1994,ant1999,rei2012b,jos2017}, as in a pseudostreamer
  configuration \citep{tor2011}. Even models possessing the same topology may
  yield different results depending on the geometry and shape of the magnetic
  field lines \cite[e.g.,][]{ste2001b,mas2013}. Discriminating between the
  characteristics of these various scenarios and understanding how the details
  of the topologies affect the evolution of an eruption requires a larger
  ensemble of events than we were able to include in the work presented here.}

To more definitively conclude that access to open field influences whether an
X-class flare is likely to be eruptive, a larger sample of non-eruptive flares
is needed.  This sample might be accomplished, for example, by relaxing the
requirement used here that the flare be sited west of central meridian, though
by doing this there is a concern that the open-field regions on the Sun may
not be accurately determined by the PFSS model. This would probably increase
the risk of a flare location being classified as having access (or non-access)
to open flux in a way that is difficult to quantify.

Alternatively, the sample might be expanded to include flares of smaller
magnitude. While including such smaller magnitude events would result in
better statistics, it would also raise the question of whether non-eruptive
flares are such because they lack the energy to fully propel a CME or whether
they are non-eruptive due to a lack of access to open field. In reality, these
two factors (energy deposited vs.\ access to open field) are likely linked,
given that a very energetic event may be able to push through a small amount
of closed field to access open field that would otherwise be inaccessible for
less energetic cases.

In this investigation, we focused on whether an X-class flare was sited in a
location with access to open fields. However, the topology of the coronal
magnetic field is complex, and contains narrow channels of open flux wedged
between closed domains of connectivity. Closed fields may lie underneath
separatrix domes associated with coronal null point, or they may be found
under the large helmet surface(s) that often wrap around the Sun. With a
larger sample size, the specific topologies associated with both eruptive and
non-eruptive flares may become more apparent.

\acknowledgements

This material is based upon work supported by the National Science Foundation
under Grant No.~1357018 to Lockheed Martin.  G.B.\ also acknowledges support
from NASA under award number NNX14AD45G.

\appendix

\section{Statistical Considerations}
\label{section:stats}

To quantitatively evaluate whether access to open field influences whether an
X-flare is eruptive, consider the following two models. In the first model,
$M_1$, access to open field does not play a role in determining whether an
X-flare is eruptive.  In the second model, $M_2$, X-class flares are eruptive
at a different rate when there is access to open field compared to when there
are no nearby open fields.  To determine which of these models is more likely,
we compute the Bayes factor (odds ratio), which is a statistic that compares
the likelihood of getting the observed data from each of the models.  More
explicitly, the two models for the observations are:
\begin{itemize}
  \item[]{$M_1$: The probability that an X-flare will produce an eruption,
    $p_e$, is independent of the proximity to open field.}
  \item[]{$M_2$: The probability that an X-flare will produce an eruption
    depends on whether there is access to open field at the flare site, where
    $p_o$ is the probability that the site of the X-class flare is located
    near access to open field, and where $p_c$ is the probability that an
    X-flare occurs in a location without any nearby open field.}
\end{itemize}

The data used to evaluate the likelihood of each of these models is summarized
in a contingency table $D$ (i.e., as shown in Table~\ref{table:contingency}),
whose elements are:
\begin{itemize}
  \item[]{$n_{eo}$: the number of eruptive X-class flares with access to open
    field.}
  \item[]{$n_{no}$: the number of non-eruptive X-class flares with access to
    open field.}
  \item[]{$n_{ec}$: the number of eruptive X-class flares under closed field.}
  \item[]{$n_{nc}$: the number of non-eruptive X-class flares under closed
    field.}
\end{itemize}

The probability of the observed contingency table $D$ resulting from each of
the models $M_i$, assuming binomial random variables, can now be calculated.
For $M_1$, the probability of getting $D$, for a given eruption probability
$p_e$, is
\begin{equation}
  Pr(D \vert p_e, M_1) = \frac{n_o! n_c!}{n_{eo}! n_{no}! n_{ec}! n_{nc}!}\,
  p_e^{n_e} (1 - p_e)^{n_n},
\end{equation}
where $n_e$ is the number of eruptive X-class flares, $n_n$ is the number of
non-eruptive X-class flares, $n_o$ is the number of X-class flares from ARs
with access to open field, and $n_c$ is the number of X-class flares from ARs
under closed field.  Marginalizing over  $p_e$ results in the
following probability of the data, assuming a uniform prior on $p_e$ given
model $M_1$ ($Pr(p_e \vert M_1)=1$ for $0 \le p_e \le 1$):
\begin{eqnarray}
  Pr(D \vert M_1) &=& \int_0^1 dp_e\, Pr(p_e \vert M_1)\, Pr(D \vert p_e,
  M_1)\\ &=& \frac{n_o! n_c!}{n_{eo}! n_{no}! n_{ec}! n_{nc}!} \frac{n_e!
    n_n!}{(n_e + n_n + 1)!}.
\end{eqnarray}

For $M_2$, the probability of getting $D$, given probabilities $p_o$ and
$p_c$, is
\begin{equation}
  Pr(D \vert p_o, p_c, M_2) = \left[\frac{n_o!}{n_{eo}! n_{no}!}\, p_o^{n_{eo}} (1 -
  p_o)^{n_{no}}\right]\left[ \frac{n_c!}{n_{ec}! n_{nc}!}\, p_c^{n_{ec}} (1 -
  p_c)^{n_{nc}}\right].
\end{equation}
Marginalizing over $p_o$ and $p_c$ results in the probability of the data,
assuming uniform priors on $p_o$ and $p_c$ given model $M_2$ ($Pr(p_o \vert
M_2)=1$ for $0 \le p_o \le 1$, $Pr(p_c \vert M_2)=1$ for $0 \le p_c \le 1$):
\begin{eqnarray}
  Pr(D \vert M_2) &=& \int_0^1 dp_o \int_0^1 dp_c \, Pr(p_o \vert M_2)\,
  Pr(p_c \vert M_2)\, Pr(D \vert p_o, p_c, M_2)\\ &=& \frac{n_o! n_c!}{(n_o +
    1)! (n_c + 1)!}.
\end{eqnarray}
Given $Pr(D \vert M_1)$ and $Pr(D \vert M_2)$, the Bayes factor $K$
is therefore given by \citep[e.g.,][]{kas1995}:
\begin{eqnarray}
  K &=& \frac{Pr(D \vert M_1)}{Pr(D \vert M_2)}\\ &=& \frac{n_e! n_n! (n_o +
    1)!  (n_c + 1)!}{n_{eo}! n_{no}! n_{ec}! n_{nc}! (n_e + n_n + 1)!},
\end{eqnarray}
where a value of $K=1$ indicates that both models are equally likely to
produce the observed contingency table, i.e., neither model is more likely
than the other with the choice of an uninformative prior \citep[see, e.g.,
  section 3.2 of][for the interpretation of the Bayes factor]{kas1995}.
% Also see table 2 of Jamil et al 2016.
For the values given in Table~\ref{table:contingency}, the Bayes factor is
$K=0.74$.

To determine how sensitive the result is to the choice of priors, we repeat
this analysis using a delta function at the Maximum Likelihood Estimate for
the values of $p_*$ for each model.  Marginalizing over the fraction $p_e$ to
get the probability of the data for model $M_1$ with $Pr(p_e \vert M_1) =
\delta(p_e - n_e/(n_e+n_n))$ gives
\begin{eqnarray}
  Pr(D \vert M_1) &=& \int_0^1 dp_e \, Pr(p_e \vert M_1)\, Pr(D \vert p_e,
  M_1)\\ &=& \frac{n_o! n_c!}{n_{eo}! n_{no}! n_{ec}! n_{nc}!} \frac{n_e^{n_e}
    n_n^{n_n}}{(n_e + n_n)^{n_e + n_n}}.
\end{eqnarray}
For model $M_2$, marginalizing over the fractions $p_o$ and $p_c$ to get the
probability of the data for model $M_2$ with $Pr(p_o \vert M_1) = \delta(p_o -
n_{eo}/n_o)$ and $Pr(p_c \vert M_1) = \delta(p_c - n_{ec}/n_c)$ gives
\begin{eqnarray}
  Pr(D \vert M_2) &=& \int_0^1 dp_o \int_0^1 dp_c \, Pr(p_o \vert M_2)\,
  Pr(p_c \vert M_2)\, Pr(D \vert p_o, p_c, M_2)\\ &=& \frac{n_o! n_c!}{n_{eo}!
    n_{no}!  n_{ec}! n_{nc}!} \frac{n_{eo}^{n_{eo}}
    n_{no}^{n_{no}}}{n_o^{n_o}} \frac{n_{ec}^{n_{ec}} n_{nc}^{n_{nc}}}{n_c^{n_c}}.
\end{eqnarray}
Thus, the Bayes factor for these priors is given by
\begin{eqnarray}
  K &=& \frac{Pr(D \vert M_1)}{Pr(D \vert M_2)}\\ &=& \frac{n_e^{n_e}
  n_n^{n_n} n_o^{n_o} n_c^{n_c}}{(n_e + n_n)^{n_e + n_n} n_{eo}^{n_{eo}}
      n_{no}^{n_{no}} n_{ec}^{n_{ec}} n_{nc}^{n_{nc}}},
\end{eqnarray}
which has a value $K=0.12$ for the values given in
Table~\ref{table:contingency}.  This value indicates that model $M_2$ is much
more likely.

The conclusion clearly depends on the choice of priors, in part because of the
extremely small number of non-eruptive flares.  The two sets of priors chosen
represent the extremes, and thus the real Bayes factor should lie between
these two. 

\facility{GOES/XRS, SDO/AIA, SDO/HMI, SOHO/EIT, SOHO/LASCO, SOHO/MDI}

%% \software goes here

\bibliographystyle{apj}

\end{document}